\documentclass[letter]{article}
 \usepackage{graphicx}
 \usepackage{amsmath}
 \usepackage{hyperref}
 \usepackage{color}

\begin{document}
 
 \begin{center}
 
 {\bf \Large Hate: no choice. Agent simulations.}\\[5mm]

 {\large Krzysztof Ku{\l}akowski, Ma{\l}gorzata J. Krawczyk \\
and Przemys{\l}aw Gawro\'nski}\\[3mm]

 {\em
 
 Faculty of Physics and Applied Computer Science, AGH University of Science and Technology, al. Mickiewicza 30, PL-30059 Krak\'ow, Poland\\

 }

 
 {\tt kulakowski@novell.ftj.agh.edu.pl}

 \bigskip
 
 \today
 
 \end{center}
 
 \begin{abstract}

We report our recent simulations on the social processes which -- in our opinion -- lie at the bottom of hate. First simulation deals with the so-called Heider balance where initial purely random preferences split the community into two mutually hostile groups. Second simulation shows that once these groups are formed, the cooperation between them is going to fail. Third simulation provides a numerical illustration of the process of biased learning; the model indicates that lack of objective information is a barrier to new information. Fourth simulation shows that in the presence of a strong conflict between communities hate is unavoidable.

 \end{abstract}
 
 \noindent
 
 
 \noindent
 

 \bigskip

\section{INTRODUCTION}

To write that hate as a subject on a physicist's desk is a challenge is an understatement. For obvious reasons, we are not able to comment what has been written on hate only recently by philosophers, psychologists, sociologists and historians. Still, having in mind what we can deduce about hate from our computer simulations, we feel that it is at least more honest and clear to confess how we imagine the role of the simulation itself in this kind of research.\\

The simulation is the part of research most safe. Not a surprise, that in various sociophysical papers we the readers are fed with computational details. The devil is before and after; in the model assumptions and in the interpretation of the results. We the authors are somewhat conscious of these dangers, but this knowledge is even not necessary to do what we actually do. Actually in most cases we meet the problems with silence. If, then, our simulations could be of interest for sociologically oriented readers, the reality described by means of these simulations must be simple -- this is a necessary condition. As a rule, the input of the model contains some numbers even if their measurement is dubious; a computational model is always expressed with numbers and we are forced to imagine that data are accessible in this or that way. The output -- the results of the simulation -- should be detectable qualitatively; only then we are able to compare with reality at least the existence of effects obtained in our computers. Summarizing, a typical result of the simulation is that in these and these conditions some effect does appear. We the authors like to think that we simulate some kind of a deterministic machine, the action of which is described with "social" variables. This kind of understanding is not far from what can be found in sociological textbooks. "A social mechanism (...) is a constellation of entities and activities that are linked to one another in such a way that they regularly bring about a particular type of outcome" (Hedstr\"om, 2005, p. 11).\\

"Humans are not numbers." Wrong; we just don't want to be treated as numbers - this statement comes from a famous sociophysicist (Stauffer, 2003), as a mark of old discussion about the unique character of human being in Nature. One of consequences of a possible consensus in this matter would be to decide to what extent the character of laws in social sciences is the same as in physics. While such a conclusion seems to be far, it seems appropriate to specify at least approximately the conditions, when we expect that human behaviour is to some extent deterministic. In our opinion, this is possible when the exceptional character of external circumstances meets standard individuals, who can be represented -- for the purposes of the simulation - by numbers. We do not expect that these individuals show any outstanding intelligence, knowledge or heroism. If me meet them on street, we are not more impressed, than some other time with other people. On the contrary, the demand of an external situation should hit basic human needs, as physiological needs, needs of safety or of belongingness (Maslow, 1950). In these conditions we can expect that individual characteristics of people becomes less important; what we learnt in our lives seems to be not applicable. These rare situations provide fuel for wide spectrum of social simulations, from the theory of rational choice (von Neumann and Morgenstern, 1944; Szab\'o and F\'ath, 2008) to human bodies as moving particles under some social forces (Helbing, 1993).\\

We want to supplement these vague considerations by two remarks. First is the natural position of our field of research between necessity and free will. A subject under full control is out of scientific interest -- everything we need is known. On the contrary, a research on subjects which are completely uncontrolled has no vital applications. What is of interest is at the border, where results of research can put forward our abilities. An example of what is at the border in human life is a social norm; after years of efforts some people are able to modify their norms to some extent, while others live and die in the same moral environment. Norms form the sociological context, and human behavior cannot be evaluated in a separation from her or his social norms. Summarizing, in our opinion social norms are at the center of any valuable sociological research. In particular, relation of individuals to hate -- whom? -- is the subject of a social norm, which can be constitutive for a community.\\

Second remark is some continuation of what is told above about standard people and non-standard situation. The main handicap of social sciences when compared to physics is the unrepeatability of the measurement. In opinion of at least one coauthor of this text, this difficulty comes more from the context than from the people. People are unique, yes, but this kind of uniqueness is less relevant in situations which we would like to simulate. The point is that it is possible to exchange actors leaving the piece untouched, but not the opposite. If historical events form a pattern, we can talk about a mechanism in the sense of Hedstr\"om, which appears more than once. This is rare, but we the researchers must assume that this is possible.\\

Hate as a socially mediated state of mind overlaps with themes which are well established in socially oriented simulations (Castellano et al, 2009): social norms, cooperation and public opinion. Their common content is a picture of a heterogeneous community, where seemingly unimportant differences can lead to a split into groups. As a consequence, cooperation and contact between members of different groups is deteriorated, mutual understanding is substituted by ignorance, finally the social labeling and hostility emerge between members of different groups. These processes lie at the bottom of hate.\\

Below we are going to report our recent simulations on these social processes. First simulation described in Section II deals with the so-called Heider balance (Ku{\l}akowski, 2007) where initial purely random preferences split the community into two mutually hostile groups. 
Three subsequent sections describe three other simulations. Second simulation (Ku{\l}akowski and Gawro\'nski, 2009) shows that once these groups are formed, the cooperation between them is going to fail. Third simulation (Kulakowski, 2009; Malarz et al, 2009) provides a numerical illustration of the process of biased learning; the model indicates that lack of objective information is a barrier to new informations. Fourth simulation (Ku{\l}akowski, 2006) is devoted to the impossibility of cooperation with enemies in the presence of a strong conflict. At the end, there is also some place for conclusions.

\section{HEIDER BALANCE}

Our story on hate begins from almost nothing. Let us consider personal contacts in some small community. For each two individuals their mutual acceptance depends on intangible details, often imperceptible for others. For some of them this acceptance is better, for some it is worse; if we measure their values with any method, we find that they evolve in time. We can refer to (Murray, 2002) where a method was presented of a measurement of the dynamics of marital interactions. As it was observed in some cases, a mutual exchange of repulsive signals can break off the tie. When such a conflict emerges in front of the community, it influences also the quality of other contacts and so on. Everybody has to decide how the event modified his or her relations with the others. As the arising tensions are often in mutual contradiction, the task is quite complex.\\

 \begin{figure}[ht]
 \centering
 {\centering \resizebox*{12cm}{9cm}{\rotatebox{-90}{\includegraphics{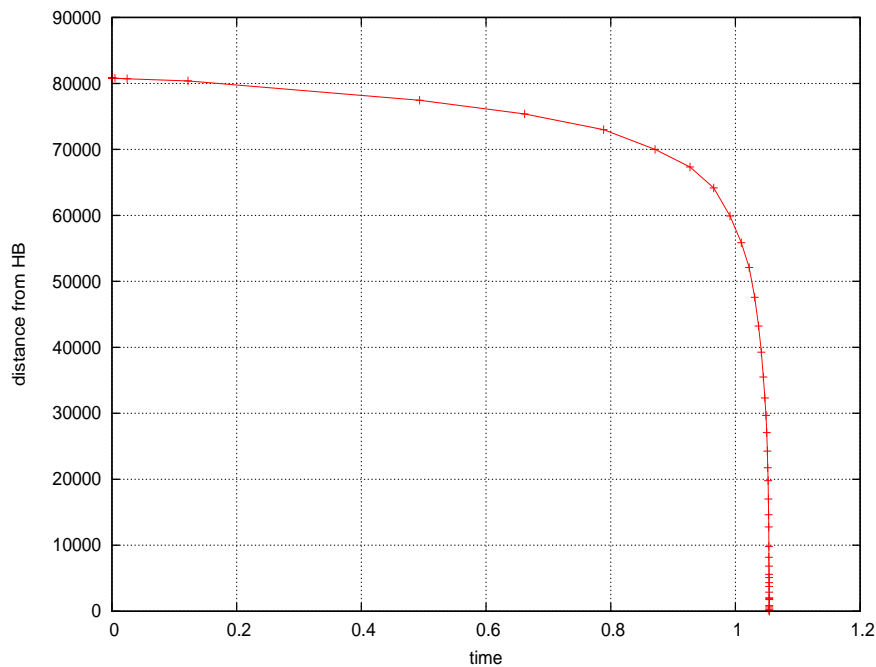}}}}
 \caption{The distance from the state of Heider balance, measured in the number of "negative" triangles, against time.}
 \label{fig-1}
 \end{figure}

Faced with such contradictory requests, people find however a way to resolve their internal conflicts. Namely, they modify some of their opinions and attitudes in such a way that their coherence is restored. This mechanism was recognized in a series of laboratory experiments by Festinger and other researchers (Festinger, 1957; Aronson, 1992) and it is known as removing the cognitive dissonance. In this way some negative ties become positive, some positive ones become negative. The final result can be somewhat surprising: the whole community becomes divided into two mutually hostile groups. Still, all the ties within each group are friendly. This state does not evolve any more, because the situation is clear: everybody knows who is enemy and who is friend. Moreover, each friend of other friend is a friend, each enemy of a friend and each friend of an enemy are enemies. This kind of equilibrium is known as the Heider balance (Heider, 1958).\\

The ideas of the cognitive dissonance and of the Heider balance found an empirical illustration in a controlled psychological experiment by Zachary (Zachary, 1977). For two years, he observed a group of 34 members of a karate club. These observations allowed to construct a friendship networks and to store it in the form of a matrix. Later, a conflict emerged between the administrator of the club and the club's instructor and the group split into two camps (Girvan and Newman, 2002).\\

Obviously, importance of the work by Zachary for social networks was recognized by sociologists (Hage, 1979; Wasserman and Faust, 1994). As described by Freeman (Freeman, 2008), new impact in sociological applications of the graph theory was given by physicists. Since the publication of the paper of Girvan and Newman (Girvan and Newman, 2002), tens of algorithms have been applied to the Zachary's data. The idea was to check if a given algorithm can reproduce the actual division of the club members. \\

Now we have to be more specific. The common aim of all these methods was to determine clusters of graph nodes, where nodes are connected more densely within the clusters than between them. A careful reader notices, that the term "more densely" is only qualitative. A mathematically-oriented reader guesses that the task is probabilistic: we set a hypothesis, that clusters do exist, and we extract these clusters from the graph. The actual data may support the hypothesis with some probability. At the end we decide if the obtained probability is large enough. If yes, we accept the obtained cluster structure. Recent reviews on the computational methods of cluster identification can be found in (Fortunato and Castellano, 2009; Fortunato, 2009).\\

The simulation which we want to underscore here was designed specifically for the problem of the cognitive dissonance. Suppose that each two nodes are connected, what means that every group member has some relations with each other. In the simplest model the links, which represent the friendly or hostile relations between people, are marked with signs: plus or minus, respectively. The cognitive dissonance appears when one negative link makes a triangle with two positive ones, or when three negative links make a triangle. The former case means that two friends of somebody are mutually hostile. The latter is that three individuals are mutually hostile. Lack of cognitive dissonance is then equivalent to a simple condition that the product of three links which make a triangle must be positive. On the contrary, the number of triangles where this product is negative can serve as a quantitative measure of the cognitive dissonance.\\

The Heider balance attracted attention of several authors -- for a review of recent computations see (Ku{\l}akowski, 2007). In our calculation (Ku{\l}akowski et al, 2005), the links are represented by real numbers $A(i,j)$ between, say, -1 and +1, where $i, j$ are the numbers of nodes. Within this range, their time evolution is governed by the differential equation

\begin{equation}
\frac{dA(i,j)}{dt}=\sum_k A(i,k)A(k,j)
\end{equation}
In other words, the relation $A(i,j)$ between individuals $i$ and $j$  improves, if most other indviduals $k$ are either friends of both or enemies of both. In these cases, the product $A(i,k)A(k,j)$ is positive. On the contrary, if many individuals $k$ are either enemies of $i$ and friends of $j$ or the opposite, the relation between $i$ and $j$ gets worse: $A(i,j)$ decreases. This is a direct mathematical realization of the removal of the cognitive dissonance. Indeed, the simulation ends in the state without "negative" triangles; this is the state of balance in the Heider sense.\\

Taking into account the continuous scale of relations allows to observe the character of the time evolution. What is surprising in the obtained data is that the number of the "negative" triangles evolves initially slowly. At a given moment, however, the rate of changes increases very abruptly; an example is presented in Fig. 1. There is yet another interesting result: our Eq. 1 with the Zachary initial data reproduces exactly the division of 34 club members into two groups (Ku{\l}akowski et al, 2005).\\

What we can learn about hate from this simulation is absurdity of its origin. Here we have a real psychological mechanism, namely removing of the cognitive distance. We apply it each day to make our emotional environment more ordered and clear. With this mechanism our hostilities directed towards casual people, formed by occasional events and sometimes irrational thoughts, are transformed into a strong, unambiguous and ordered prejudice against a well defined group. Everybody whom we respect, i.e. each member of our group, confirms that this prejudice is justified. In our group, the group of respected people, this prejudice is a social norm.\\

What is next?

\section{IN-GROUP COOPERATION}

In this section we examine one of the currently investigated mechanisms of cooperation, namely the competitive altruism (Roberts, 1998). We show that this mechanism is not immune for the limitation of cooperation to the own player's group. This means, that cooperation does not protect against hate; we can just cooperate with some and hate others.\\

 \begin{figure}[ht]
 \centering
 {\centering \resizebox*{12cm}{9cm}{\rotatebox{-90}{\includegraphics{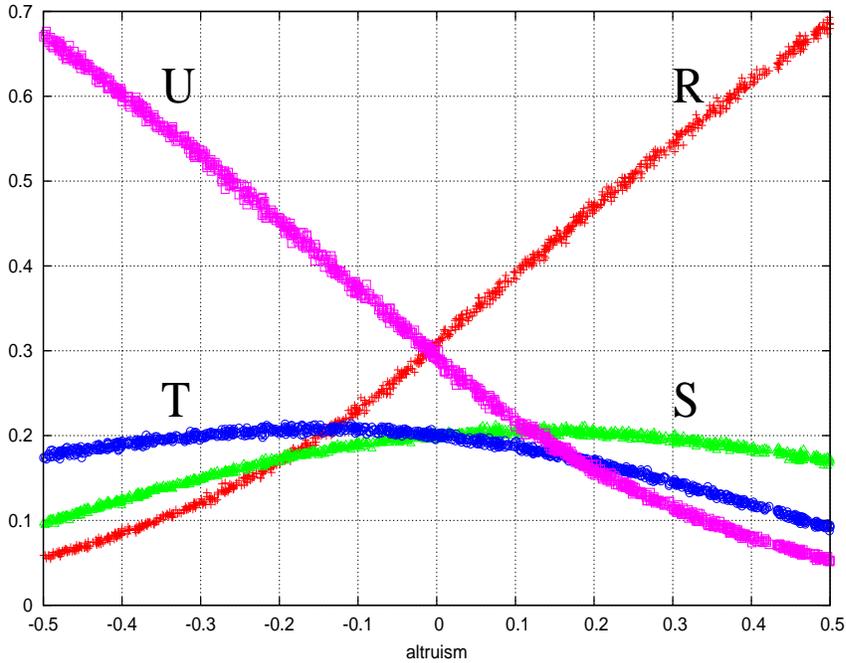}}}}
 \caption{The probability to be in position R, S, T and U against the altruism level $e$ of the player.}
 \label{fig-2}
 \end{figure}

The problem of cooperation has been discussed usually in terms of the Prisoner's Dilemma, abbreviated as PD. In some sense, this dilemma is a challenge for our culture, as it demonstrates its clash with the individual rationality. There is a story which justifies the term PD; as it is commonly known, we prefer another example. Two individuals are imprisoned, and the food is provided between 3.00 AM and 6 AM. A loyal strategy is to get up near, say, 7.00 AM , and to share food evenly. However, the portions are scanty and it is tempting to get up earlier and eat more. The dilemma is to sleep till 7.00 AM or not; once both lie in wait for food, they don't sleep well. In an adopted terminology about PD, to sleep peacefully means "to cooperate", and to lie in wait means "to defect". The amount of situations which could illustrate PD is innumerable, from the Kyoto Protocol back to the biblical story on the Prodigal Son.\\

Standard representation of PD in game theory is the payoff matrix (Straffin, 1993)\\

\begin{tabular}{|r|c|l|}

\hline
 & Cooperate & Defect \\
\hline
Cooperate & R, R & S, T \\
Defect	& T, S	& U, U \\
\hline
\end{tabular}
\bigskip

When a cooperator meets a cooperator, the payoff is R (reward) for both. When both of them defect, both get U (uncooperative). However, once a cooperator meets a defector, the latter gets T (temptation), and the former gets S (sucker's payoff). The game can be classified as PD if $S < U < R < T$. Additional condition is that $T+S < 2R$, to exclude the cyclic strategy when one cooperate and other defects or the opposite; this strategy is not PD. As $R < T$ and $S < U$, to defect is always better than to cooperate; then, according to game theory, a rational player should defect. The point is that if both defect, they get less than if both cooperate.\\

 \begin{figure}[ht]
 \centering
 {\centering \resizebox*{12cm}{9cm}{\rotatebox{-90}{\includegraphics{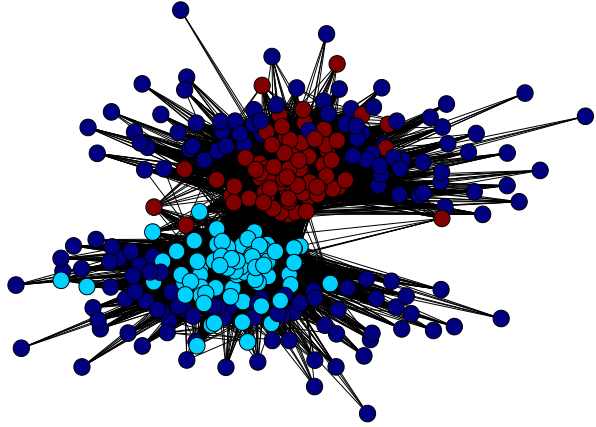}}}}
 \caption{Graph visualization of strong cooperation within two groups, with players with negative altruism left aside; $\kappa=0.3$}
 \label{fig-3}
 \end{figure}

As we see, PD is designed as to disable cooperation of rational players. Still we can imagine that people cooperate for reasons other than a simple individual profit. These reasons can be kinship and direct or indirect reciprocity; for recent reviews on cooperation theory see for example (Fehr and Fischbacher, 2005). Motivation by kinship is limited to the own group. Motivation by reciprocity can also lead to in-group favoritism, if the group members expect reciprocation from the same group (Fehr and Fischbacher, 2005). Before we switch to the motivation by the competitive altruism, let us quote one social experiment on randomly formed groups. \\

The aim of this research (Goette et al, 2006) was to check if the mere dependence to a group causes a non-selfish behaviour between its members. The PD experiment with some modifications was conducted in the frames of the Swiss army, where the assignment of individuals to platoons was completely random. Still, after three-weeks training, the cooperation between members of the same platoon was significantly higher, than between members of different platoons. We note that hostility between different platoons has not been observed -- fortunately hate does not appear so easily. \\

We refer to this experiment as it provides a demonstration, that the formation of a group or a community is equivalent to the formation of some social norm, which unifies the group. This equivalence was clearly expressed by a historian, Elias Bickerman: "The first need of any social system is to create incentives to make people do more work than that required by their immediate wants." (Bickerman, 1973, p.73). This means also that expectation of cooperation of members of the same group is stronger. In longer time scale, such an expectation becomes trust. \\

The theory of competitive altruism (Roberts, 1998; Van Vugt et al, 2007) explores the cooperation as a demonstration of resources. First assumption is that individuals differ in their level of altruism. Second is that people compete to establish altruistic reputation. In accordance of the Zahavi handicap principle (Zahavi and Zahavi, 1997), the signal must be costful to be reliable. The altruistic behaviour fulfils this condition. Our question is, if this mechanism can be responsible for the group formation, with enhancement of cooperation within groups. Then we could draw a line from the Goette's experiment to the concept of competitive altruism. The problem of hate is left aside till the end of this section.\\

To address the above question, we resort to a simulation. The model is adopted from our previous work (Ku{\l}akowski and Gawro\'nski, 2009); having the paper published we found that its assumptions can be justified within the theory of competitive altruism. These assumptions are: \\

\noindent
1.	Agents $i$ differ in their altruism level $e(i)$.\\
2.	The probability $P(i,j)$ that an agent $i$ cooperates with another agent $j$ increases with the altruism of $i$ and with the reputation of $j$. Namely, we used the sum $P(i,j)=e(i)+W(i)$, limited to the range (0,1).\\
3.	The reputation $W(i)$ of $i$ increases when $i$ cooperates, and decreases when $i$ defects. \\

Most important result of the model is shown in Fig. 2. As we see, the probability that $i$-th agent meets a cooperative co-player depends on the altruism of $i$. Most altruistic agents experience cooperation in seventy percent of games. This is true for any payoff we would like to assign to particular outcomes of PD. Then, the values of the payoffs are not relevant.\\

To investigate the case of two groups, the scheme of simulation was modified (Gawro\'nski et al, 2009) by adding a constant $\kappa$ to the probability of cooperation if the co-player belongs to the same group; in the opposite case, the same constant $\kappa$ was subtracted. This modification altered the matrix of the probability of cooperation $P(i,j)$. If the constant $\kappa$ is large enough, this matrix shows a cluster structure; the method is described in (Krawczyk, 2008). Namely, two clusters could be detected: agents in the same clusters cooperate more frequently, than agents in different clusters. A half of agents is left besides the clusters; still, they are more connected either to one or to the other group. This structure is visualized in Fig. 3.\\

In this way we made a second step on our path. First was the prejudice. Second is the loss of trust in members of group or community other, than the agent's own group. 

\section{BIASED LEARNING}

Third step to hate is the lack of understanding. \\

At the beginning of Section II the method was remarked of a measurement of dynamics of marital status (Murray, 2001). The method was to distinguish couples of low and high risk level by controlling the signals they send during a conversation. These signals, coded into a special system, were gathered by Gottman and Levenson (Gottman and Levenson, 1992) for 73 couples in 1983. The couples were investigated again in 1987. Among the high risk couples, the percentage of divorced was 19 against 7 in low risk group. Then, the character of conversation allowed to predict the future. \\

In the case of political differences, the outcome can be even more explicit. Character of anonymous statements about politics in the Internet was investigated by Sobkowicz and Sobkowicz (Sobkowicz and Sobkowicz, 2009). In many cases, comments of internauts could be classified as an exchange of insults, with increasing hostility. The point is more than often, the discussion was dominated by adherents of two major Polish political parties, both Center-Right: PiS and PO. The analysis shown that to be hated by opponents is the desired goal as well as to be admired by supporters. What is the origin of this extreme behaviour? We are going to look for the answer in our third simulation, which deals with the public opinion.\\

The motivation for the simulation comes from the Zaller theory on the mass opinion (Zaller, 1992). It is based on four ideas: "The first is that citizens vary in their habitual attention to politics and hence in their exposure to political information and argumentation in the media. The second is that people are able to react critically to the arguments they encounter only to the extent that they are knowledgeable about political affairs. The third is that citizens do not typically carry around in their heads fixed attitudes on every issue on which a pollster may happen to inquire; rather, they construct "opinion statements" on the fly as they confront each new issue. The fourth is that, in constructing their opinion statements, people make greatest use of ideas that are, for one reason or another, most immediately salient to them..." (Zaller, 1992).\\

Keeping -- as we believe -- the content of these ideas, we reformulated the mathematical scheme of the simulation (Ku{\l}akowski, 2009). Further generalization of our formalism (Ku{\l}akowski and Gronek, 2009; Malarz et al, 2009) is close to the Deffuant model of public opinion (Deffuant et al, 2000). Therefore we term the formulation as the Zaller-Deffuant model from now on. Its core is as follows. Individuals are the subject of a stream of messages produced by the mass media. An individual is represented by a set of messages which she/he received in the past. Further, a message is represented as a point on the plane of issues. The condition to receive a new message is that its distance from any previously accepted message should be shorter than some critical distance $a$. This critical distance is a measure of the mental capacity of an individual. Each individual starts at it first message with position selected randomly at the plane of issues. Once a new message appears, it is received or not by an individuals, depending on her/his history. The communication between the individuals is maintained by messages, formulated and sent by the individuals. The position of such a message is determined by a temporal average of the messages received previously by the sender.\\

The output of the simulation is the distribution of probability p that an opinion formulated by an individual on a given issue is YES. For each individual, this probability can be calculated as\\

\begin{equation}
p=\frac{\sum_t x_t \Theta(x_t)}{\sum_t |x_t|}
\end{equation}
where $\Theta(x)=0$ for $x<0$, $\Theta(x)=1$ for $x>0$, and the sums are performed over messages accepted at time $t$. In particular, $p$ is equal to unity (YES for sure) if all messages received by the individual are on the positive side of the OY axis. On the contrary, $p = 0$ (NO for sure) if the $x$-th coordinate of all these messages is negative. The obtained distribution of $p$ is calculated over all individuals.\\

In the simulation, the system as a whole is the subject of a specific test. The stream of  messages from the media is symmetric with respect to the axis OY. Then, the individuals have no more arguments for YES than for NO and the opposite. Smart individuals should recognize this symmetry and their probabilities $p$ should be close to 1.  Indeed, if the critical distance $a$ is large enough, the distribution of $p$ has a maximum close to $p=1/2$.  However, for small $a$ the distribution of $p$ shows a bimodal character: two maxima appear, for $p$ negative and $p$ positive (Fig. 4). This means, that small mental capacity leads to extreme opinions (Ku{\l}akowski, 2009). In other words, less clever people are more sure of their opinions. This paradoxical result can be seen as a sociological counterpart of the physical concept of the spontaneous symmetry breaking. \\

 \begin{figure}[ht]
 \centering
 {\centering \resizebox*{12cm}{9cm}{\rotatebox{-90}{\includegraphics{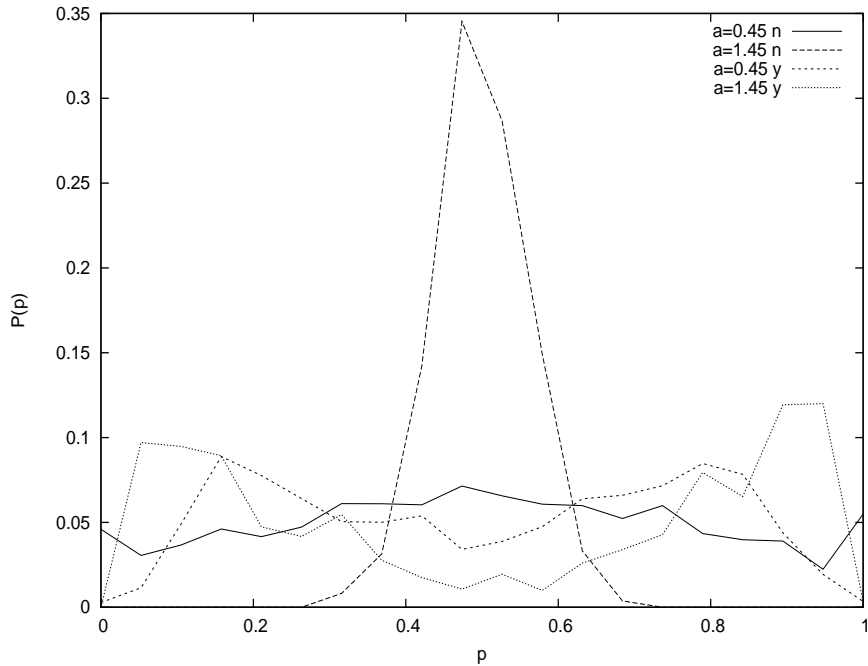}}}}
 \caption{The probability distribution $P(p)$ of the probability $p$, that an individual opinion about the model issue is YES. Out of four plots,
the only one centralized near $p=0.5$ is the plot for the mental capacity $a=1.45$, without interaction (marked by $n$). Other plots are for $a=0.45$ with and without interaction, and for $a=1.45$ with interaction ($y$).}
 \label{fig-4}
 \end{figure}

Here we are interested in the consequences of the interpersonal communication. It appears that this form of interaction deteriorates the central maximum of the distribution $P(p)$, as shown also in Fig. 4 (Ku{\l}akowski and Gronek, 2009; Malarz et al, 2009). This indicates, that the communication between the group members can maintain prejudice. Such a conclusion coincides with the warnings against the groupthink (Janis, 1972). Even if individuals are endowed with high capacity, the exchange of messages leads to a formation of one single opinion, which is an average over some finite set. The stream of incoming messages is dominated by those exchanged within the system. As a consequence, all individual opinions are the same. The community becomes perfectly homogeneous.\\

Suppose now that this community contacts with another one, with different opinion. Till this moment, neither mutual understanding nor tolerance was necessary. Individual differences were hidden to maintain the social coherence -- now they are exposed. If hostility appears, we should not be surprised.
 
\section{CASE OF GHETTO}

To get hate, the last missing ingredient is a spark. It can be imperceptibly small: a discussion between shepherds, a soldier's joke, a tabloid article. Real wars do not break out like that, real wars are provoked by serious people, who want to be sure on the results of their actions. A good provocation is a murder of a respected person, as Mohandas Ghandi, Martin King, Icchak Rabin. We do not need to discuss an initiation -- this is easy. We are interested in the question, if -- in the presence of full conflict - the development of hate can be spontaneously halted. \\

The subject of our last simulation is a ghetto. Historically, the term "ghetto" is ambiguous. Here we use it to describe the situation where the position of people is defined by two traits: i) an attempt of an inhabitant of ghetto to leave the area makes his situation worse, ii) human laws, as understood by inhabitants, are broken by an external power (Ku{\l}akowski, 2006). The decision is limited to two strategies: to support the resistance or to obey the rules imposed by the external power. In the psychological dimension, the choice is to hate or not to hate. \\

In our considerations, we refer to the Maslow theory (Maslow, 1950) remarked at the introduction. According to this theory, we fulfil our needs in a definite order. The physiological needs are most basic and they must be fulfilled at first. Once they are satisfied, we start to bother about safety. Next is the needs for belongingness, esteem and self-actualization, in this order. Our point is that vital decisions of people are biased by their needs. This is in accordance of the fourth postulate of Zaller: "in constructing their opinion statements, people make greatest use of ideas that are, for one reason or another, most immediately salient to them" (Zaller, 1992), quoted in the preceding section. In other words, the decision of resistance and hate is taken in connection of the actual need, yet not satisfied.\\

Even in a ghetto, people differ in their position and possibilities. There are some, who struggle to satisfy their physiological needs: water and food. These efforts makes their behavior determined and predictable. They have no opportunity to express their hate -- this is too costly. Actually, the simplest way to reduce the resistance of people is to reduce their life to this physiological level. This method is well known and still being applied (Politkovskaya, 2003). There are some other people, who are relatively safe; they can bother about their belongingness, love, friendship and perhaps the higher needs . Their decisions about resistance depend on their individual social entanglements; an attempt to simulate this would be idle. Who remains are the people who concentrate on their safety. Their decision about resistance is connected to this need. Will they be more safe, when obeying the rules of the external power? This is the question most salient for them. \\

The simulation consists in the numerical solution of the so-called Master equation (van Kampen, 1981)\\

\begin{equation}
\frac{dr}{dt}=-u(r)r+w(r)(1-r)
\end{equation}
where $r$ is the time dependent probability of the choice of resistance and hate, $u(r)$ is the probability of changing the decision from resistance to submission, $w(r)$ is the probability of changing the decision from submission to resistance. What is sought after is the function $r(t)$, while the functions $u(r)$ and $w(r)$ should be given. We note that on the contrary to standard application of this equation in physics of magnetism, $w(r)$ and $u(r)$ are not expected to be equal for $r=1/2$. The shape of these functions is difficult to be determined, but we can state two details: {\it i)} $w(r)$ increases with $r$, and {\it ii)} $u(r)$ decreases with $r$. These facts follow directly from a simple intuition: once everybody resists, it is difficult to submit; once submission is common, conversions to resistance are rare. In physics we speak on a positive feedback.\\

To demonstrate the consequences, let us select the simplest choice of two linear functions: $u(r)= a(1-r)$, $w(r)=br$. Then there are two constant solutions: $r = 0$ (total submission) and $r = 1$ (total resistance). First solution is stable if $a > b$, second one is stable if $b > a$. In this simple case, the time-dependent solution is available\\

\begin{equation}
r(t)=\frac{r_0}{r_0+(1-r_0)\exp[(a-b)t]}
\end{equation}
where $r_0$ is the initial value of $r$.  As we see, the result is determined by the sign of the expression $a-b$. If it is positive, the submission prevails. If it is negative, the resistance is common. \\

 \begin{figure}[ht]
 \centering
 {\centering \resizebox*{12cm}{9cm}{\rotatebox{-90}{\includegraphics{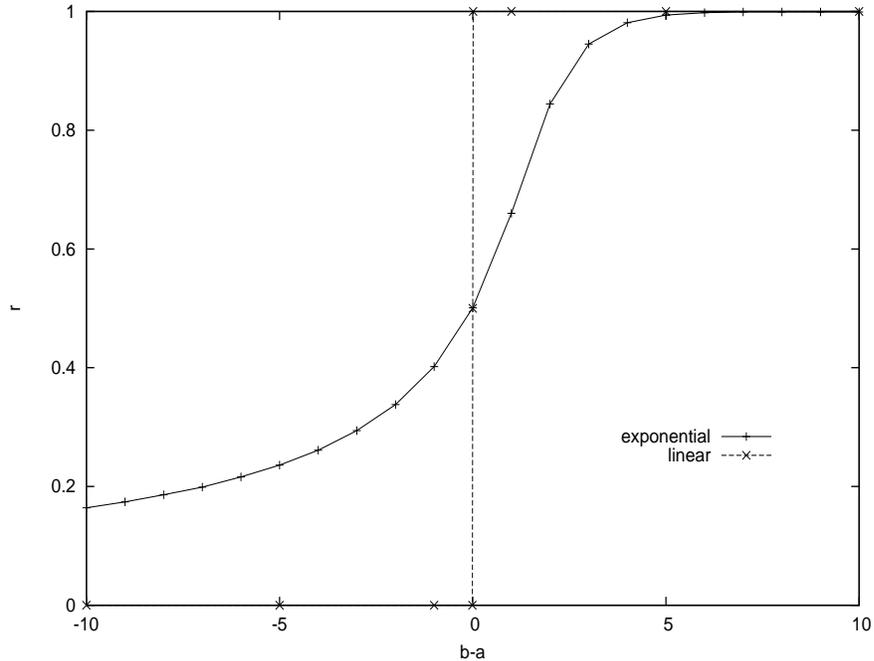}}}}
 \caption{Stationary values of the probability $r$ against the difference $b-a$ for linear and exponential forms of the functions $u(r)$ and $w(r)$. 
The exponential functions are chosen after (Ku{\l}akowski, 2006).}
 \label{fig-5}
 \end{figure}

We should add two remarks. First is that the functions $u(r)$ and $w(r)$ remain unknown. Moreover, our propositions of their linear shape can be treated as useful approximations only within some range of $p$. However, the qualitative character of the solution is preserved also for another choice of these functions -- see the comparison in Fig. 5. Second remark appeals to the political reality. Once an individual chose the resistance, his safety will be not improved when he decides back to submit, as -- sooner or later -- he will be called to account by the external power. That is why it is reasonable to expect that $a$ is always less than $b$. For the external power, there are two strategies: either to declare an amnesty to improve the safety of ex-opponents, or to delay the problem, keeping both constants $a$ and $b$ as small as possible. However, if $b$ is already large, the second strategy is not possible and the first strategy is ineffective.\\

At the beginning of this section we asked the question, if -- in the presence of full conflict - the development of hate can be spontaneously halted. Our results indicate, that the answer is no. 

\section{CONCLUSION}

We demonstrated four schemes of simulations, intended to reflect some features of real sociological mechanisms. Three of them - prejudice, distrust and misunderstanding towards members of other communities - can contribute to the development of hate. Although they seem necessary, they are not sufficient conditions of this process. On the other hand, they are sometimes functional for a given community. If this is the case, the community develops social norms to support prejudice, distrust and misunderstanding, and these norms are obeyed in particular by numerous good, healthy and agreeable individuals. \\

The last calculation treats with the case when the hate is already present. The society is perhaps unified against the external power, perhaps split. In both cases to hate enemies is a duty and an identification. This kind of hate unifies the community; people have no choice to hate or not to hate, they have only the choice to hate whom. A demonstration of this kind of hate is equivalent to a demonstration of friendship to members of the own community. On the contrary to a private hate against a given person, this kind of hate is not destructive for the psyche. This kind of hate does not preclude friendship, sacrifice and love; it just excludes the members of hated communities. \\

It is not our intention here to analyze how this kind of hate turned out to be destructive for human race, even if it is still functional for some communities. Today we are educated enough to know that it is possible to live without hate; then the human nature only enables hate, but does not imply it. Therefore, the research on hate should be directed more to communities and their social norms, than to rare twists of individual minds.\\

\bigskip

{\bf REFERENCES}\\

\noindent
Aronson, E. (1992). The Social Animal. New York: Freeman.\\
Bickerman, E. J. (1972). The Ancient Near East, in J.A. Garraty and P. Gay (Eds.) The Columbia History of the World. New York: Harper and Row Publishers.\\
Castellano, C., Fortunato, S., Loreto, V. (2009). Statistical physics of social dynamics. Reviews of Modern Physics 81, 591-646.\\
Deffuant, G., Neau, D., Amblard, F., Weisbuch, G. (2000). Mixing beliefs among interacting agents.  Adv. Complex Systems 3, 87-98. \\
Fehr, E., Fischbacher, U. (2005). Human altruism -- proximate patterns and evolutionary origins. Analyse und Kritik, 27, 6-47. \\
Festinger, L. (1957).  A theory of cognitive dissonance. Stanford: Stanford UP.\\
Fortunato, F., Castellano, C. (2009). In R. A. Meyers (Ed.) Encyclopedia of Complexity and Systems Science, Vol. 1. Berlin: Springer. \\
Fortunato, S. (2009). Community detection in graphs, arXiv:0906.0612 \\
Freeman, L. C. (2008). Going the Wrong Way on a One-Way Street: Centrality in Physics and Biology. Journal of Social Structure, 9, No 2.\\
Gawro\'nski, P., Krawczyk, M. J., Ku{\l}akowski, K. (2009). Altruism and reputation: cooperation in groups, arXiv: 0903.3902.\\
Girvan, M., Newman, M. (2002). Community structure in social and biological networks, Proc. Natl. Acad. Sci. USA 99, 7821-7826.\\
Goette, L., Huffman, D., Meier, S. (2006). The impact of group membership on cooperation and norm enforcement: Evidence using random assignment to real social groups, working paper No. 06-7, Federal Reserve Bank of Boston.\\
Gottman, J. M., Levenson, R. W. (1992). Marital processes predictive of latter dissolution : behavior, psychology and health. J. Personality and Social Psychol., 633, 221-233.\\
Hage, P. (1979). Graph Theory as a Structural Model in Cultural Anthropology, Annual Review of Anthropology, 8, 115-136.\\
Hedstr\"om, P. (2005). Dissecting the Social: On the Principles of Analytical Sociology, Cambridge: Cambridge UP.\\
Heider, F. (1958). The Psychology of Interpersonal Relations.  New York: Wiley and Sons.\\
Helbing, D. (1993). Boltzmann-like and Boltzmann-Fokker-Planck equations as a foundation of behavioral models. Physica A, 196, 546-573.\\
Janis, I. L. (1972). Victims of Groupthink : A psychological study of foreign-policy decisions and fiascoes. Boston: Houghton Mifflin Comp. \\
Van Kampen, N. G. (1981). Stochastic Processes in Physics and Chemistry. Amsterdam: Elsevier.\\
Krawczyk M. J. (2008). Differential equations as a tool for community identification. Phys. Rev. E, 77, 065701(R) 1-4.\\
Ku{\l}akowski, K. (2007).  Some recent attempts to simulate the Heider balance problem. Computing in Science and Engineering, 9, 80-85.\\
Ku{\l}akowski, K. (2009). Opinion polarization in the Receipt-Accept-Sample model. Physica A, 388, 469-476.\\
Ku{\l}akowski, K. (2006). Cooperation and defection in ghetto, Int. J. Mod. Phys. C 17, 287-298.\\
Ku{\l}akowski, K., Gawro\'nski, P. (2009). To cooperate or to defect? Altruism and reputation. Physica A, 388, 3581-3584.\\
Ku{\l}akowski, K., Gawro\'nski, P., Gronek, P. (2005). The Heider balance -- a continuous approach.  Int. J. Mod. Phys. C, 16, 707-716. \\
Ku{\l}akowski, K., Gronek, P. (2009). The Zaller-Deffuant model of mass opinion, presented at the Dynamics Days Europe 2009, G\"ottingen, Aug 31 - Sept 4, 2009. \\
Malarz, K., Gronek, P., Ku{\l}akowski, K. (2009). The Zaller-Deffuant model of public opinion, arXiv:0908.2519. \\
Maslow, A. H. (1954). Motivation and Personality. New York: Harper.\\
Murray, J. D. (2001). Mathematical Biology I: An Introduction. New York: Springer.\\
von Neumann, J., Morgenstern, O. (1944).  Theory of Games and Economic Behavior. Princeton: Princeton UP.\\
Politkovskaya, A. (2003). A Small Corner of Hell. Dispatches from Chechnya. Chicago: Univ. of Chicago Press.\\
Roberts, G. (1998). Competitive altruism: from reciprocity to the handicap principle. Proc. R. Soc. Lond. B 265, 427-431.\\
Sobkowicz, P., Sobkowicz, A. (2009). Dynamics of hate based networks. arXiv:0905.3751.\\
Stauffer, D. (2004). Introduction to Statistical Physics outside Physics. Physica A, 336, 1-5.\\
Straffin, P. D. (1993). Game Theory and Strategy. Washington, D.C.: Math. Association of America.\\
Szab\'o, G., F\'ath, G. (2008). Evolutionary games on graphs. Physics Reports 446, 97-216.\\
Van Vugt, M., Roberts, G., Hardy, C. (2007). Competitive altruism: Development of reputation-based cooperation in groups, in R. Dunbar, L. Barrett, (Eds.) Handbook of Evolutionary Psychology. Oxford: Oxford University Press.\\
Wasserman, S., Faust, K. (1994). Social Network Analysis: Methods and Applications. Cambridge: Cambridge UP.\\
Zachary, W. W. (1977).  An information flow model for conflict and fission in small groups, Journal of Anthropological Research, 33, 452-473.\\
Zahavi, A., Zahavi, A. (1997). The Handicap Principle: The Missing Piece of Darwin's Puzzle. Oxford: Oxford University Press.\\
Zaller, J. R. (1992). The Nature and Origins of Mass Opinion. Cambridge: Cambridge University Press.\\

 \end{document}